# Probing the Phonon Mean Free Paths in Dislocation Core by Molecular Dynamics Simulation

Yandong Sun[a,e], Yanguang Zhou[b], Ming Hu[d], G. Jeffrey Snyder[e,*], Ben Xu[c,*], and Wei Liu[a,*],

[a]Laboratory of Advanced Materials, School of Materials Science and Engineering, Tsinghua University, Beijing 100084, People's Republic of China.

[b]Department of Mechanical and Aerospace Engineering, The Hong Kong University of Science and Technology, Hong Kong, People's Republic of China.

[c]Graduate School, China Academy of Engineering Physics, Beijing 100193, People's Republic of China.

[d]Department of Mechanical Engineering, University of South Carolina, Columbia, SC 29208, USA.

[e]Department of Materials Science and Engineering, Northwestern University, Evanston, IL 60208, USA.

*Email:

jeff.snyder@northwestern.edu

bxu@gscaep.ac.cn

weiliu@mail.tsinghua.edu.cn

**Abstract**

Thermal management is extremely important for designing high-performance devices. The lattice thermal conductivity of materials is strongly dependent on the structural defects at different length scales, particularly point defects like vacancies, line defects like dislocations, and planar defects such as grain boundaries. Traditionally, the McKelvey-Shockley phonon Boltzmann's transport equation (BTE) method combined with molecular dynamics simulations has been widely used to evaluate the phonon mean free paths (MFPs) in defective systems. However, this method can only provide the aggregate MFPs of the whole sample. It is, therefore, challenging to extract the MFPs in the different regions with different thermal properties. In this study, the 1D McKelvey-Shockley phonon BTE method was extended to model inhomogeneous materials, where the effect of defects on the phonon MFPs is explicitly obtained. Then, the method was used to study the phonon interactions with the core structure of an edge dislocation. The phonon MFPs in the dislocation core were obtained and consistent with the analytical model such that high frequency phonons are likely to be scattered in this area. This method not only advances the knowledge of phonon-dislocation scattering but also shows the potential to investigate phonon transport behaviors in more complicated materials.

## 1. Introduction

Phonons are the primary heat carriers in materials, especially in semiconductors[1]. Enriching the knowledge of phonon transport is important for managing the lattice thermal conductivity ($\kappa_L$) of materials to achieve better device performance. The Boltzmann transport equation (BTE)[2], Green's function[3], Landauer-Büttiker formalism[4], and molecular dynamics (MD)[5] simulations have been developed to model phonon transport in materials. By combining the force constants and eigenvectors from *ab initio* calculations, the BTE, and Green's function can well describe phonon transport in thermally homogeneous materials. Defects have been verified experimentally to effectively decrease the $\kappa_L$[6]. Point defects, dislocations, interfaces, precipitates, and nanograins combine to scatter phonons of different wavelengths[7] to achieve the lowest $\kappa_L$. However, *ab initio* calculations are unsuitable for or incapable of obtaining realistic phonon transport in crystals with defects owing to their small computational size. Defects modeled in *ab initio* calculations usually have a very high concentration (e.g., point defects) or neglect some features of the defect (e.g., the long distance nature of strain field induced by dislocations). The computational scale of MD simulations is more than $10^3$ times larger than that of *ab initio* calculations; thus, MD is more suitable for simulating phonon transport in crystals with defects.

Non-equilibrium MD (NEMD) simulations have been widely used to study phonon transport in pure crystals, in crystals with point defects, dislocations, and interfaces, and even in amorphous materials[8]. The $\kappa_L$ can be directly obtained from a NEMD simulation, but detailed phonon scattering information is missing. To overcome this shortcoming, the frequency-domain direct decomposition method was developed to calculate the spectral heat flux in NEMD simulations[9,10], from which the contributions of different phonons to the total heat flux can be obtained. The heat flux can also be modeled using the 1D McKelvey-Shockley phonon BTE method[11], and the phonon mean free paths (MFPs) can be calculated using the length-dependent heat flux results[12]. However, the traditional method of fitting the phonon MFPs only applies to thermally homogeneous materials, which assumes that the phonon scattering rates are constant throughout the material. Therefore, it is necessary to extend the method to thermally inhomogeneous materials, where the phonon scattering rate varies in the regions containing defects.

In this study, the 1D McKelvey-Shockley flux method was extended to thermal inhomogeneous materials and then it was used to study the phonon scattering with dislocation core. A line of atoms in the PbTe crystal was deleted to simulate the core structure of the edge dislocation. Thermal inhomogeneities were caused near the position of the dislocation core. Then a heat flux perpendicular to the dislocation line was added in the NEMD simulation. The $\kappa_L$ and spectral heat flux was directly obtained. The $\kappa_L$ of the dislocation core structure model decreased by 19% compared to the pristine structure model. By using the extended 1D McKelvey-Shockley flux method, the frequency range of the phonons scattered by the dislocation and the corresponding phonon MFPs due to phonon-dislocation scattering were obtained.

## 2. Methodology

### 2.1 NEMD Simulation Setup

The LAMMPS[13] package was used to conduct the NEMD simulations with a timestep of 5 fs. A Buckingham pair potential taken from ref.[14] was used to model the interactions between atoms. A total of $4.5 \times 10^6$ steps were performed, corresponding to a running time of 28 ns. The first 12.5 ns was used to relax the structure with the isothermal-isobaric ensemble (NPT) and the canonical ensemble (NVT), then the following 15 ns was used to obtain a steady temperature gradient and heat flux. Langevin thermostats were applied to the hot and cold region with a damping parameter of 0.1 (20 fold of the timestep of the simulation) and keyword tally "yes." See the schematic of simulation cell in Fig. 1. The main system region was coupled to the microcanonical ensemble (NVE). The velocity of atoms for spectral heat flux calculation and the temperature distribution along the heat flux direction were sampled in the last 1.5 ns.

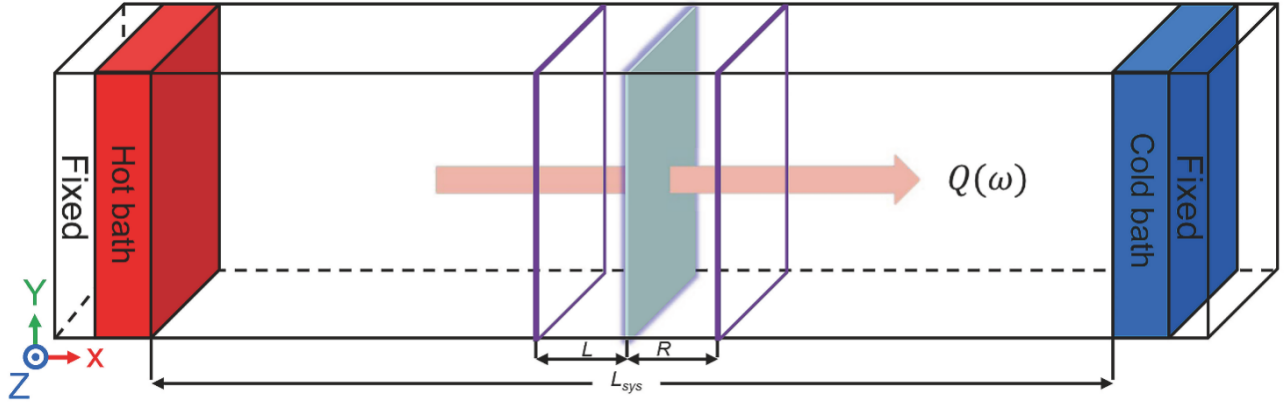

Fig. 1 Schematic of the simulation cell used in the NEMD simulations. A virtual interface at the position of the dislocation core of the model was selected, and the velocities of the atoms in the left and right regions of the virtual interface (within 2 nm) were sampled at successive timesteps to calculate the spectral heat flux, Q, through the interface.

## 2.2 Spectral Heat Flux Method

The spectral heat flux ($Q(\omega)$) was calculated using the following equation[9,10]

$$Q(\omega) = -\frac{2}{t_{simu}\omega A}\sum_{i\in L, j\in R}\sum_{\alpha,\beta\in\{x,y,z\}}\text{Im}\langle \tilde{v}_i^\alpha(\omega)^* K_{ij}^{\alpha\beta} \tilde{v}_j^\beta(\omega)\rangle , \qquad (1)$$

where $t_{simu}$ is the total simulation time, $A$ is the area of the cross-section, α and β are the Cartesian directions, and $K_{ij}^{\alpha\beta}$ is the second-order interatomic force constant between the atoms distributed on both sides of the dislocation (denoted by $L$ and $R$), $\tilde{v}_i^\alpha(\omega)$ and $\tilde{v}_j^\beta(\omega)$ are the discrete Fourier-transform velocity of atom $i$ in direction α and atom $j$ in direction β, and $\tilde{v}_i^\alpha(\omega)^*$ is the complex conjugate of $\tilde{v}_i^\alpha(\omega)$; the angular brackets denote the steady-state nonequilibrium ensemble average, Im denotes the imaginary part of a complex number. The velocity of atoms was directly obtained from the NEMD simulations, and the interatomic force constants were obtained by the finite displacement method in MD simulations.

## 2.3 Extension of One-Dimensional McKelvey-Shockley Flux Method with Defects

It is possible to use the McKelvey-Shockley phonon BTE method to describe phonon transport in NEMD simulations[11]. A schematic of the McKelvey-Shockley flux method is shown in Fig. 2. Steady-state 1D transport along $x$ with an infinite $y$-$z$ plane was assumed. The forward/backward phonon fluxes ($F^+(x)$ and $F^-(x+dx)$, respectively) incident on a slab with thickness $dx$ transmit or reflect with the backscattering probabilities per length, $\frac{1}{\lambda(\omega)}$. Here $\lambda(\omega)$ is the spectral phonon MFP for backscattering defined as the average distance travelled by a carrier before backscattering that contributes to the outward fluxes $F^-(x)$ and $F^+(x+dx)$. The equations are

$$\frac{dF^+(x,\omega)}{dx} = -\frac{F^+(x,\omega)}{\lambda(\omega)} + \frac{F^-(x,\omega)}{\lambda(\omega)} \tag{2}$$

$$-\frac{dF^-(x,\omega)}{dx} = +\frac{F^+(x,\omega)}{\lambda(\omega)} - \frac{F^-(x,\omega)}{\lambda(\omega)}. \tag{3}$$

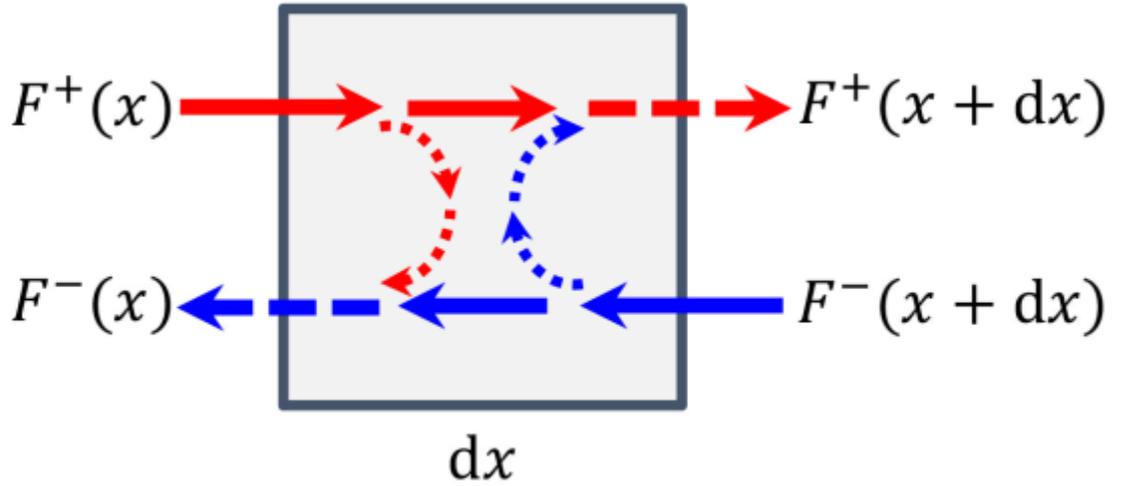

Fig. 2 Schematic of the McKelvey-Shockley flux method. The forward/backward phonon fluxes ($F^+(x)$ and $F^-(x+dx)$, respectively) which are incident on a slab with thickness $dx$ transmit or reflect with the backward scattering probabilities per length $\frac{1}{\lambda(\omega)}$.

By specifying the incoming fluxes at the boundaries, the forward and backward phonon fluxes can be obtained by solving Eqs. (2) and (3). The boundary conditions are as follows,

$$F^+(x = 0, \omega) = F_B(\omega, T_H) \tag{4}$$

$$F^-(x = L, \omega) = F_B(\omega, T_C), \tag{5}$$

where $L$ is the length of the model, and $F_B(\omega, T_H)$ and $F_B(\omega, T_C)$ are the incoming phonon fluxes from the hot and cold baths, respectively, (Fig. 1; the interface contact resistance is neglected here). We will first assume a fully homogeneous system. The solutions are

$$F^+(x, \omega) = F_B(\omega, T_H) - \frac{F(\omega)}{\lambda(\omega)} x \tag{6}$$

$$F^-(x, \omega) = F_B(\omega, T_C) - \frac{F(\omega)}{\lambda(\omega)}(x - L), \tag{7}$$

where the net phonon flux $F(\omega) = F^+(x, \omega) - F^-(x, \omega)$, which is constant in the steady-state. The forward and backward phonon fluxes vary linearly along the heat flux direction with a slope of $-\frac{F(\omega)}{\lambda(\omega)}$.

Then, $F(\omega)$ can be extracted by subtracting Eq. (7) from Eq. (6) and isolating $F(\omega)$,

$$F(\omega) = \frac{1}{1+L/\lambda(\omega)}[F_B(\omega, T_H) - F_B(\omega, T_C)]. \tag{8}$$

The spectral heat flux ($Q(\omega)$) corresponds to multiplying the net phonon flux ($F(\omega)$) by the phonon energy ($\hbar\omega$),

$$Q(\omega) = \frac{1}{1+L/\lambda(\omega)} \hbar\omega * [\ F_B(\omega, T_H) - F_B(\omega, T_C)] = \frac{Q_0(\omega)}{1+L/\lambda(\omega)}, \tag{9}$$

which can be interpreted as the Landauer-Büttiker flux ($Q_0(\omega)$) with an effective transmission function[15]

$$T_{eff}(\omega) = \frac{1}{1+L/\lambda(\omega)}. \tag{10}$$

The above expression for $Q(\omega)$ can be applied to all length scales and spans from the ballistic to diffusive transport regimes. In the ballistic limit, $L \ll \lambda$ and $T_{eff} \to 1$, which indicates that all the phonons can achieve ballistic transport through the model without scattering. The population of phonons in the system will be constant; thus, $Q(\omega)$ is independent of the length. In the diffusive limit, $L \gg \lambda$ and $T_{eff} \approx \lambda/L$, which indicates that the population of phonons will change linearly in the system. Therefore, $Q(\omega) \propto 1/L$, which is expected from Fourier's law. When $L \approx \lambda$, part of the phonons exhibit ballistic transport, while others exhibit diffusive transport.

If the transmission functions of models with different lengths are calculated, $\lambda(\omega)$ can be obtained by fitting the data to Eq. (10). Alternatively, the spectral heat flux of the model with $L_1$ length can be used as a reference, and the ratio between the model with $L_1$ length and the model with $L$ length can be calculated[16] as,

$$R(\omega, L) = \frac{Q(\omega, L_1)}{Q(\omega, L)} = 1 + \frac{L - L_1}{\lambda(\omega) + L_1}, \tag{11}$$

where $L_1$ is a constant which is decided by the reference model, $R(\omega, L)$ is a linear function of $L$ and $\lambda(\omega)$ can be obtained by fitting the slope in a defect-free, homogeneous system.

However, the interpretation becomes more complex when there is a defect or inhomogeneity in the model because of the different phonon scattering rates and mean free paths in different parts of the system. The equations for the 1D McKelvey-Shockley phonon BTE must then be rewritten as a piecewise function, where the averaged backscattering probability per length for two different areas are noted as $\frac{1}{\lambda_d(\omega)}$ and $\frac{1}{\lambda_p(\omega)}$, for the defect region $x_1$ to $x_2$ (area I) and the pristine, defect-free region outside of it (area II), respectively. The equations for $F^+(x, \omega)$ are then expressed as

$$\frac{dF^+(x,\omega)}{dx} = -\frac{F^+(x,\omega)}{\lambda_p(\omega)} + \frac{F^-(x,\omega)}{\lambda_p(\omega)}, \quad x < x_1, \ x > x_2, \tag{12}$$

$$\frac{dF^+(x,\omega)}{dx} = -\frac{F^+(x,\omega)}{\lambda_d(\omega)} + \frac{F^-(x,\omega)}{\lambda_d(\omega)}, \quad x_1 < x < x_2, \tag{13}$$

and the same equations can define $F^-(x, \omega)$.

Under the same boundary conditions as in Eqs. (4) and (5), the following can be obtained,

$$F(\omega) = \frac{1}{1 + L/\lambda_p(\omega) + d(1/\lambda_d(\omega) - 1/\lambda_p(\omega))} [F_B(\omega, T_H) - F_B(\omega, T_C)], \tag{14}$$

where $d$ is the length of area I ($d = x_2 - x_1$) which can be obtained from the temperature profile by determining the length of the abnormality in the temperature gradient.

The transmission function is

$$T'_{eff}(\omega) = \frac{1}{1+L/\lambda_p(\omega)+d(1/\lambda_d(\omega)-1/\lambda_p(\omega))}, \tag{15}$$

and the spectral heat flux ratio of the inhomogeneous $L$ model is

$$R'(\omega, L) = \frac{Q(\omega, L_1)}{Q(\omega, L)} = 1 + \frac{L - L_1}{\lambda_p(\omega) + L_1 + d(\lambda_p(\omega)/\lambda_d(\omega) - 1)}. \tag{16}$$

By comparing to Eq. (11), the above equation can be written as

$$R'(\omega, L) = \frac{Q(\omega, L_1)}{Q(\omega, L)} = 1 + \frac{L - L_1}{\lambda_{nm}(\omega) + L_1}. \tag{17}$$

Therefore, by varying the length of the inhomogeneous system, $\lambda_{nm}(\omega) = \lambda_p(\omega) + d(\lambda_p(\omega)/\lambda_d(\omega) - 1)$ can be obtained as a whole, while it is not possible to obtain $\lambda_p(\omega)$ and $\lambda_d(\omega)$ separately. It must be emphasized that $\lambda_{nm}(\omega)$ does not give the actual phonon MFPs of backscattering for area I, instead, its value implies the difference of phonon MFPs between area I and II as follows. When $d=0$, $\lambda_{nm}(\omega) = \lambda_p(\omega)$ (i.e. the homogeneous system); while $\lambda_{nm}(\omega) > \lambda_p(\omega)$ when there is a finite defect region $d$ with a suppressed mean free path (i.e. $\lambda_d(\omega) < \lambda_p(\omega)$). In most cases, $\lambda_p(\omega)$ can be fit from the pristine model, and $\lambda_{nm}(\omega)$ can be fit from the inhomogeneous system, allowing one to extract a value for $\lambda_d(\omega)$ using the definition of the nominal MFP above. This extended method can provide additional information on phonon scattering from defects in thermally inhomogeneous crystals.

## 3. Models

The above method was used to study the phonon scattering from the edge dislocation core. The dislocation core was represented as a vacancy line perpendicular to the heat flux direction in the model. The simulation was 63.15 nm long in the x-direction and had a 9.48 nm × 8.94 nm cross-section with the orientation orientation $\bar{u}_x = [\bar{1}\ 1\ 0]$, $\bar{u}_y = [\bar{1}\ \bar{1}\ 2]$, and $\bar{u}_z = [1\ 1\ 1]$. A line of an equal number of Pb and Te atoms along the z-direction was removed from the center of the model. A schematic of the simulation cell used for the NEMD simulations is shown in Fig. 1. The atoms at both ends of the system were fixed. Next to the fixed region, atoms within the length $L_{bath} = 2$ nm of the left and right side were coupled to Langevin thermostats. The temperature was fixed to 40 K in the hot bath and 0.1 K in the cold bath. The low temperature reduces the intrinsic phonon-phonon process, which helps to reveal the phonon-dislocation interactions. In classical MD simulations, all the phonon modes are populated even at extremely low system temperature. Even though the phonon distribution will change due to the quantum effect at low temperature, our work focuses on the basic physics of phonon-defect scattering rather than obtaining quantitative values for the thermal conductivity of defective materials. A virtual interface at the position of the dislocation core was chosen to calculate the spectral heat flux. Two groups of atoms (denoted by L and R) were selected on both sides of the interface, and their velocities were sampled at successive timesteps, which were used to calculate the spectral heat flux. The same simulation procedure was performed for the pristine structure.

## 4. Results and Discussion

### 4.1 Temperature Distribution and Thermal Conductivity

The temperature distributions along the heat flux direction in the dislocation core structure DC Model and pristine structure P Model are shown in Fig. 3. There was a perfect linear temperature distribution in the thermally homogeneous P Model. In contrast, there was a sharp drop in temperature at the dislocation core in the DC Model. The dislocation core area contained thermal inhomogeneities (such as increased phonon scattering); thus, a higher temperature gradient was necessary to maintain the same heat flux as that in other parts of the system. Temperature drops also occurred near the baths, which is expected when some phonon MFPs are comparable to the length of the model, such that they demonstrate ballistic transport through the sample without any scattering. The shape of the temperature profiles can be well interpreted by the McKelvey-Shockley phonon BTE method. The ballistic thermal phonons will not change the phonon population in the system, thus generating a flat temperature distribution. When combined with the linear temperature distribution due to diffusive phonons, a distribution with temperature drops near the baths and linear behavior elsewhere results.

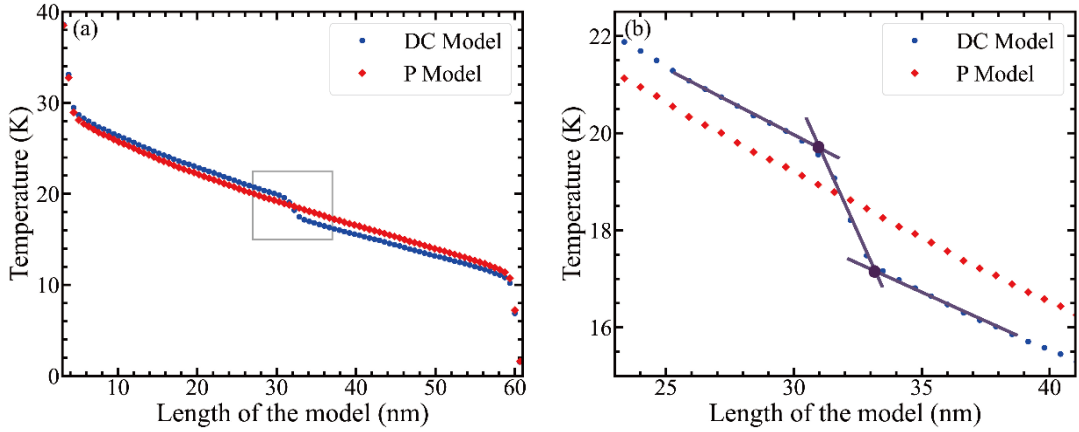

Fig. 3 (a) Temperature distributions of the dislocation core structure DC Model and pristine structure P Model in the NEMD simulations. (b) Magnified view showing of the sharp reduction in temperature near the line of vacancies (grey rectangle region in (a)).

The thermal conductivity $\kappa_L$ was calculated directly from Fourier's law, and the results are shown in Tab. 1. The temperature gradient was calculated over the entire region rather than the dislocation core area in the DC Model. The $\kappa_L$ decreased by 19.2% due to the addition of the dislocation core.

Table 1. The $\kappa_L$ calculated by Fourier's law from the NEMD simulations.

|  | Heat Flux, $J_Q$ (GW/m$^2$) | Temperature Gradient, $\partial T / \partial x$ (K/nm) | Thermal Conductivity, $\kappa_L$ (W/(m·K)) |
|---|---|---|---|
| DC Model | 4.30 | 0.331 | 13.0 |
| P Model | 4.57 | 0.287 | 15.9 |
| DC/P Ratio | 94.1% | - | 81.8% |

**4.2 Spectral Heat Flux**

Before the newly developed 1D McKelvey-Shockley, flux method can be applied to calculate the phonon MFPs, the spectral heat flux $Q(\omega)$ through the two models must be investigated first. $Q(\omega)$ is related to the thermal conductivity of the material, and describes the amount of heat transferred by phonons at frequency $\omega$ (see Fig. 4(a)). The heat flux in the dislocation core structure (DC Model) is lower in the frequency range of 1 to 2 THz compared with the pristine structure (P model), particularly at the two peaks in spectral heat flux, which indicates that these frequency range phonons were influenced by the dislocation core. Normalized spectral heat fluxes $q(\omega)$ were obtained by dividing by the total heat flux (shown in Fig. 4(b)) such that the proportion of the heat flux contribution from different frequency phonons can be determined. $q(\omega)$ did not change for frequency less than 1.6 THz, where its shape comes strictly from the phonon density of state (DOS). This is because the phonon MFP is greater than the simulation length in this frequency range, so these phonons transport ballistically without scattering. The proportional contribution of different frequency phonons to the heat flux were almost identical in the DC Model and P Model except for a small decrease in the peak at the 1.60-1.80 THz frequency range and a small increase at high frequencies. From Eq. (9), the change in $Q(\omega)$ is related to the phonon MFPs $\lambda(\omega)$ and the phonon group velocity $v(\omega)$. Thus, the phonon MFPs must be obtained.

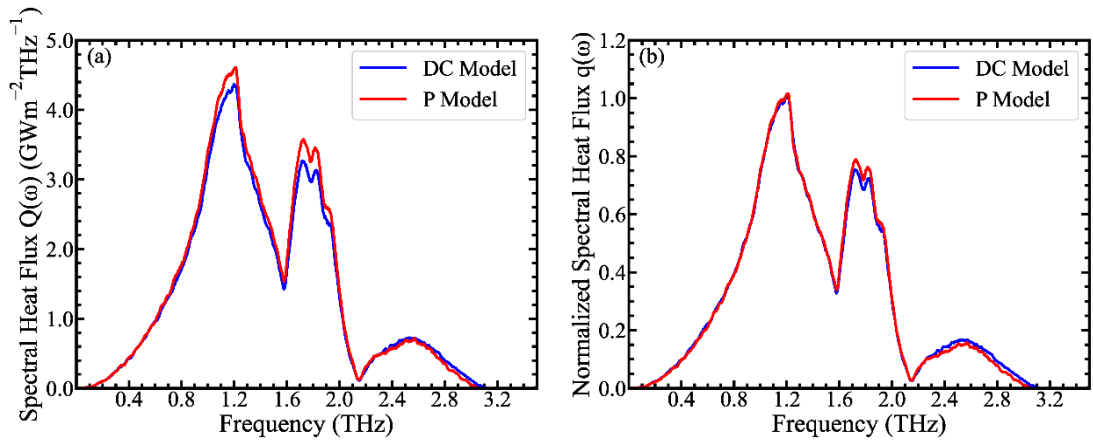

Fig. 4 (a) Spectral heat fluxes $Q(\omega)$ of the dislocation core structure DC Model and pristine structure P Model, from which the contributions from different phonons to the total heat flux can be determined. (b) Normalized spectral heat fluxes $q(\omega)$ obtained from (a) by dividing by the total heat flux, from which the proportions of the heat flux contributed by different frequency phonons can be determined. All the curves are smoothed with a width $\Delta\omega = 0.0244$ THz.

**4.3 Fitting the Phonon MFPs**

The phonon MFPs can be obtained from the length-dependent heat flux[9,17]. As mentioned previously, there are two areas in the dislocation core model DC Model. In area I with dislocation core, there is phonon-phonon scattering and phonon-dislocation core scattering. In area II outside the dislocation core, there is only phonon-phonon scattering, which is identical to the scattering process in the pristine model P Model. According to Matthiessen's rule, the backward scattering probabilities per length in area I ($1/\lambda_d$) satisfy

$$\frac{1}{\lambda_d(\omega)} = \frac{1}{\lambda_p(\omega)} + \frac{1}{\lambda_{dc}(\omega)} \tag{18}$$

where $\lambda_p(\omega)$ and $\lambda_{dc}(\omega)$ are the phonon MFPs for backward scattering from phonon-phonon scattering (sole effect in P model) and phonon-dislocation core scattering, respectively. Combining Eqs. (17) and (18) yields

$$\lambda_{nm}(\omega) = \lambda_P(\omega) + \frac{d \cdot \lambda_P(\omega)}{\lambda_{DC}(\omega)}. \tag{19}$$

The $d$ is the length of area I and can be obtained from the temperature distribution in Fig. 3 (b). It starts from the position where temperature distribution begins to deviate from the linear distribution and ends in the position where temperature distribution comes returns to the linear distribution (the distance in the horizontal direction between the two purple solid circles Figure 3b).

The nominal phonon MFPs for backward scattering $\lambda_{nm}(\omega)$ in the dislocation core structure DC Model and the phonon MFPs for backward scattering $\lambda_p(\omega)$ in the pristine structure P Model are shown in Fig. 5 (a). The length-dependent spectral heat flux results are shown in Fig. 6. According to the analysis in the previous section, if the phonons are scattered in the area with defects, $\lambda_{nm}(\omega)$ will be larger than the $\lambda_p(\omega)$ of the P Model. It was observed that $\lambda_{nm}(\omega)$ of the DC Model visibly increased when the frequency was higher than 2.1 THz, which indicates that, in this frequency range, phonons were significantly scattered by the dislocation core. By substituting $\lambda_p(\omega)$ from the P Model and $\lambda_{nm}(\omega)$ from the DC Model into the Eq. (19), the phonon MFPs for backscattering from phonon-dislocation core scattering $\lambda_{dc}(\omega)$ can be obtained. Here, only the average phonon MFP results were used to rather than the full confidence interval, and the negative values were omitted, so it was a very rough result especially in the low frequency range. The regular phonon MFPs (i.e. average distance travelled between scattering events) were 75% smaller than the phonon MFPs for backscattering[18], as shown in Fig. 5(b). The result is consistent with the analytical theory[19] that $\lambda_{dc}(\omega) \sim \omega^{-3}$ for phonon dislocation core scattering, but a constant group velocity is being assumed here and the relationship is strongly related to the spacing between the dislocation cores[20]. Most of the high frequency phonons were scattered by the dislocation core, but their contributions to the total heat flux increased according to the analysis in section 4.2. This is because there was a larger temperature gradient at the position of the dislocation core in the DC Model. It was also found that the contribution from the medium frequency phonons decreased when they were scattered less by the dislocation core. From Eq. (9), the change in $Q(\omega)$ is not only related to the phonon MFPs ($\lambda(\omega)$) but also to the phonon group velocity ($v(\omega)$). Therefore, it is necessary to check the change in the phonon group velocity after introducing the dislocation core.

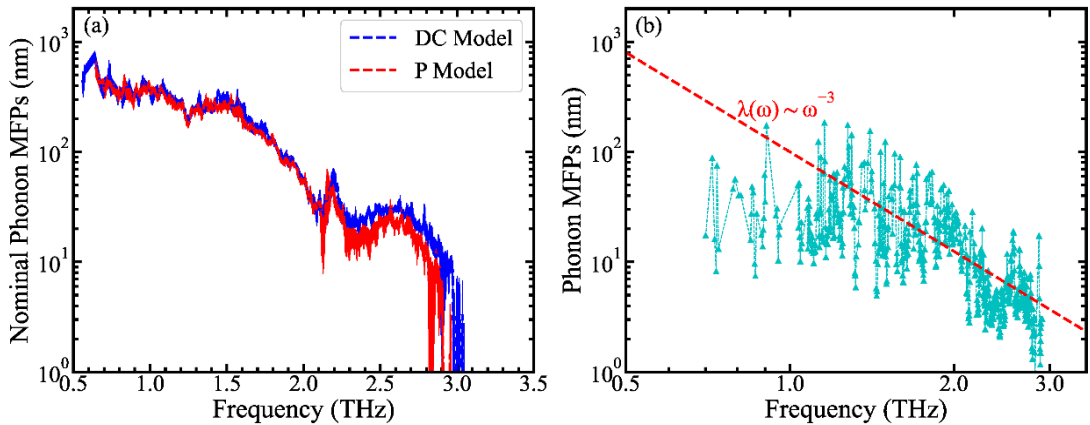

Fig. 5 (a) Nominal phonon MFPs $\lambda_{nm}(\omega)$ for the dislocation core structure DC Model and phonon MFPs $\lambda(\omega)$ for pristine structure P Model with 95% confidence intervals from the fitting processes mentioned in the Methodology section. Of note, the $\lambda_{nm}(\omega)$ is not a real phonon MFPs, a bigger value doesn't mean a weak phonon scattering, but its variations to the $\lambda(\omega)$ can reflect the phonon scattering processes in the dislocation core. High frequency phonons are significantly scattered by the dislocation core. (b) Phonon MFPs introduced by phonon dislocation core scattering assuming Mathiessen's rule, Eq 18. Here, only the average results in (a) were used to calculate the phonon MFPs instead of the whole data in confidence intervals, and the negative values were omitted.

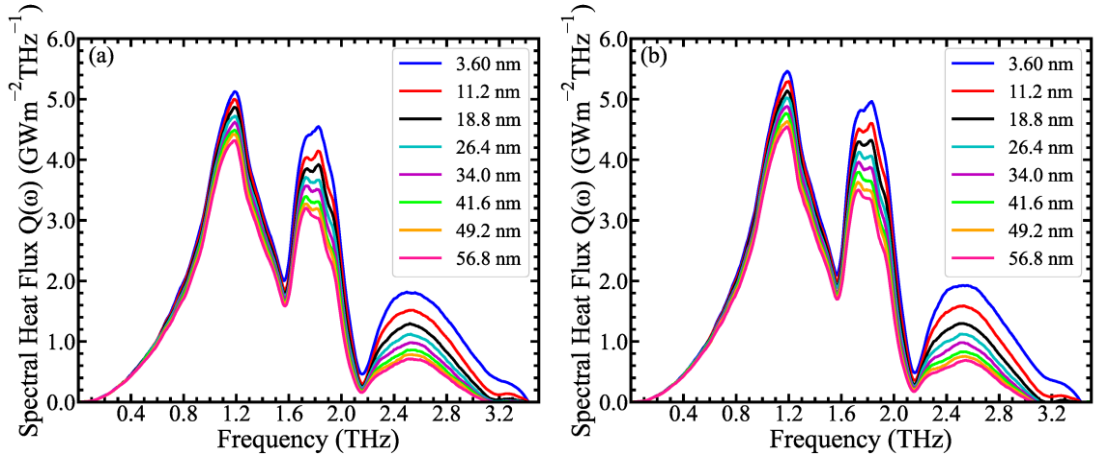

Fig. 6 Spectral heat fluxes of different length models: (a) the dislocation core structure DC Model and (b) pristine structure P Model.

**4.4 Change in the Group Velocity**

Defects and strain might also reduce the phonon velocity which has been correlated with much of the reduction in thermal conductivity observed in nanostructured PbTe materials[20]. To investigate whether a change in phonon velocity also occurs, a model with 8 × 8 × 4 unit cells in the x, y, and z directions (lattice vectors of the unit cell), respectively, was built, and a line of atoms along the z-direction in the center of the model was removed to generate the dislocation core. A lattice dynamics calculation was conducted using Phonopy[21] software, and the force constants were calculated from MD simulations with the same atomic potential that was used for the previous simulations. The group velocities of the dislocation core structure model and the corresponding pristine structure model are shown in Fig. 7. The group velocity of medium frequency phonons was greatly suppressed owing to the dislocation core and the strain it caused. Thus, the amount of heat transferred by medium frequency phonons decreased even though they were not significantly scattered by the dislocation core. This is why the contribution from phonon in the frequency range 1.0-1.6 in the D Model decrease in Fig. 4(a).

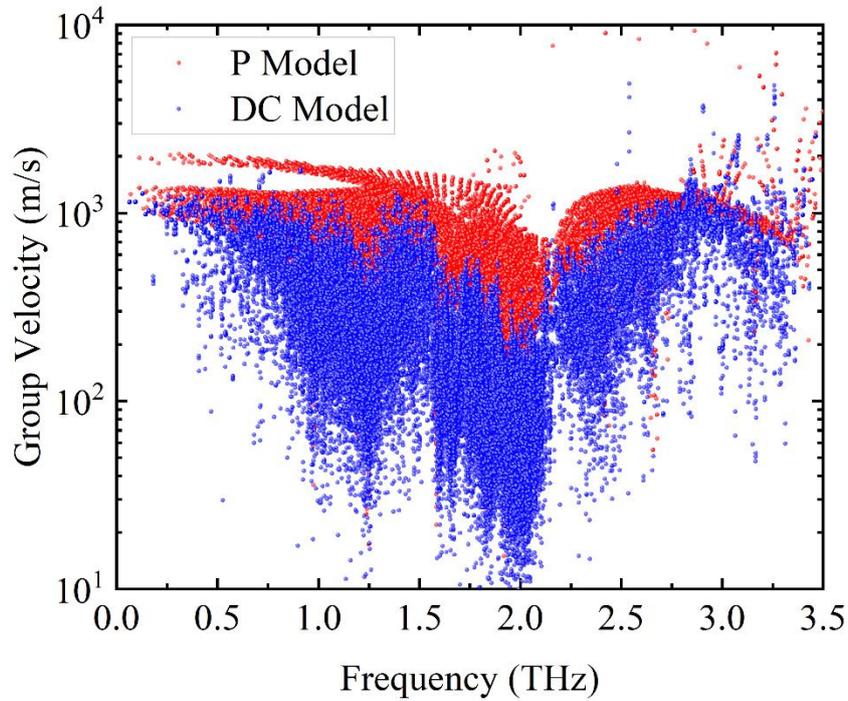

Fig. 7 Group velocities of the phonons of the dislocation core structure DC Model and pristine structure P Model from lattice dynamic calculations. The group velocity of the medium frequency phonons was greatly suppressed because of the dislocation core.

## 5. Conclusion

In this study, the McKelvey-Shockley phonon BTE method was extended to inhomogeneous materials with defects, and the phonon MFPs from phonon defect scattering were obtained. In inhomogeneous materials, it was argued that the MFP had contributions from both the homogenous and inhomogeneous (defect-containing) regions. The method was used to study the phonon scattering near the dislocation core. The variations in the phonon group velocity primarily lead to the thermal conductivity decrease. The frequency dependence of the dislocation core scattering MFP was determined, which was consistent with the analytical theory that high frequency phonons are more likely scattered by the dislocation core. Based on these results, this method illuminates the scattering mechanism of phonon transport in materials with structural inhomogeneities. The method not only advances the knowledge of phonon dislocation scattering but also shows the potential to investigate materials with more complicated components and phonon behaviors.

**Data availability**

The datasets generated during and/or analyzed during this study are available from the corresponding author on reasonable request.

Code availability

The codes used in this study are available from the corresponding author on reasonable request.

**Acknowledgment**

The authors thank Ramya Gurunathan of Northwestern University for helpful discussion and Dr. Shiyao Liu for polishing the language. Simulations were performed with computing resources granted by the National Supercomputer Center in Tianjin under project TianHe-1(A) and Quest High-Performance Computing Cluster at Northwestern University. The research reported in this publication was supported by the NSFC under grant No. 52072209, Basic Science Center Project of NSFC under grant No.51788104 and by the Tsinghua National Laboratory for Information Science and Technology. G. Jeff Snyder acknowledges support from award 70NANB19H005 from U.S. Department of Commerce, National Institute of Standards and Technology as part of the Center for Hierarchical Materials Design (CHiMaD). Ming Hu acknowledges support from the NSF (award number 2030128). Yandong Sun acknowledges support from Tsinghua University Short-term Overseas Exchange Fund.

**Appendix:**

**Author contributions**

Y.D.S. performed MD simulations, data processing, result analysis, and manuscript writing. Y.G.Z., M.H., B.X., W.L., and J. S. participated in the discussion and interpretation of results and contributed to editing the manuscript.

**Competing interests**

The authors declare no competing interests.